\documentclass[fleqn,10pt]{wlscirep}
\usepackage[utf8]{inputenc}
\usepackage[T1]{fontenc}

\usepackage{graphicx}
\usepackage{dcolumn}
\usepackage{amsmath}
\usepackage{epsfig}
\usepackage{bm}
\usepackage{amssymb}
\usepackage{hyperref}

\usepackage{algorithm} 
\usepackage{algpseudocode}

\hypersetup{colorlinks=true,linkcolor=blue,citecolor=blue,urlcolor=blue}

\title{Estimation of pure quantum states in high dimension at the limit of quantum accuracy through complex optimization and statistical inference}

\author[1,2]{Leonardo Zambrano} 
\author[1,2]{Luciano Pereira}
\author[1,3]{Sebasti\'an Niklitschek}
\author[1,2*]{Aldo Delgado}
\affil[1]{Instituto Milenio de Investigaci\'on en \'Optica, Universidad de Concepci\'on, Concepci\'on, Chile}
\affil[2]{Facultad de Ciencias F\'isicas y Matem\'aticas, Departamento de F\'isica, Universidad de Concepci\'on, Concepci\'on, Chile}
\affil[3]{Facultad de Ciencias F\'isicas y Matem\'aticas, Departamento de Estad\'istica, Universidad de Concepci\'on, Concepci\'on, Chile}

\affil[*]{aldelgado@udec.cl}

\begin{abstract}
Quantum tomography has become a key tool for the assessment of quantum states, processes, and devices. This drives the search for tomographic methods that achieve greater accuracy. In the case of mixed states of a single 2-dimensional quantum system adaptive methods have been recently introduced that achieve the theoretical accuracy limit deduced by Hayashi and Gill and Massar. However, accurate estimation of higher-dimensional quantum states remains poorly understood. 
This is mainly due to the existence of incompatible observables, which makes multiparameter estimation difficult. Here we present an adaptive tomographic method and show through numerical simulations that, after a few iterations, it is asymptotically approaching the fundamental Gill-Massar lower bound for the estimation accuracy of pure quantum states in high dimension. The method is based on a combination of stochastic optimization on the field of the complex numbers and statistical inference, exceeds the accuracy of any mixed-state tomographic method, and can be demonstrated with current experimental capabilities. The proposed method may lead to new developments in quantum metrology.
\end{abstract}
\begin{document}

\flushbottom
\maketitle
%
%
\thispagestyle{empty}


\section*{Introduction}

The basic principles of quantum mechanics strongly limit the processes that can be performed in nature. The linearity of quantum operations prohibits the possibility of creating in a deterministic way perfect copies of unknown quantum states \cite{Wootters-Zurek}. Information encoded in non-orthogonal quantum states can be perfectly recovered only at the expense of allowing a probability of error \cite{Ivanovic,Dieks,Peres0}. Measurements, a fundamental tool for understanding the natural world, are also limited. In remarkable difference with the classical realm, 
a measurement in a quantum system alters the quantum state of the system, which prevents further information gain. In this way, an ensemble of many equally prepared systems is necessary to obtain information about a quantum system and, consequently, quantum measurements become statistical in nature \cite{vonNeumann,Holevo,Peres}. Furthermore, quantum mechanics establishes through the quantum Cram\'er-Rao bound \cite{Helstrom,Helstrom2,Braunstein} a fundamental limit in the best accuracy that a measurement or estimation process can achieve as a function of the ensemble size.

The accuracy limit also bounds the estimation of unknown quantum processes and states. These are central problems in the theory of quantum measurements \cite{Paris} and play a key role in the control of quantum systems, the benchmarking of quantum technologies \cite{DiVincenzo,Kok,Nielsen}, and quantum metrology \cite{Giovannetti}. Today there is a large collection of methods \cite{Paris,Banaszek,Wootters2,James,Renes,Gross,Cramer,Ahn,Salazar,Paiva,Martinez} for estimating unknown quantum states, which are collectively known as quantum tomographic methods. These are based on the post-processing of data acquired through the measurement of a set of positive operator-valued measures in an ensemble of $N$ identically, independently prepared copies of the unknown state to be estimated. Recently, there has been a great deal of research activity aimed at designing tomographic methods with increasing accuracy, in particular methods that saturate the quantum accuracy limit. From the theoretical and experimental point of view, it has been shown that a two-stage adaptive tomographic method achieves the quantum accuracy limit in the case of unknown mixed states of a single qubit \cite{Mahler,Guo}. Unfortunately, this result does not hold in the higher dimensional case \cite{Pereira}.

Here, we report an adaptive tomographic method that asymptotically approaches the quantum accuracy limit in the important case of estimating unknown pure quantum states in high dimensions. This is an instance of multi-parameter estimation, where due to the information trade-off among incompatible observables the progress has been slow. In the case of pure states the quantum accuracy is limited by the Gill-Massar lower bound \cite{Gill-Massar}. The method is based on the concatenation of the Complex simultaneous perturbation stochastic approximation (CSPSA), a recently proposed iterative stochastic optimization method on the field of the complex numbers \cite{We}, and Maximum likelihood estimation (MLE), a well known statistical inference method \cite{Hradil2}. The method here proposed reaches the quantum accuracy limit after a small number of iterations, typically of the order of 8, for all inspected dimensions. Thereby, the method makes an optimal use of the ensemble size and surpasses the estimation accuracy of known methods for pure-state tomography \cite{Goyeneche,Carmeli,Sosa,Zambrano}. Moreover, the method also surpasses the estimation accuracy of any tomographic method designed to estimate mixed states via separable measurements on the ensemble of equally prepared copies.

\section*{Method}

The accuracy achieved in the estimation of the parameters of a quantum state can be studied by means of three fundamental inequalities. These are the Cram\'er-Rao inequality \cite{Paris2} $\mathcal{C}\leq \mathcal{I}^{-1}$, the quantum Cram\'er-Rao inequality \cite{Cramer-Rao,Paris2} $\mathcal{C}\leq \mathcal{J} ^{-1}$, and the Gill-Massar inequality \cite{Gill-Massar} $Tr(\mathcal{IJ}^{-1})\leq d-1$, where $d, \mathcal{C}$, $\mathcal{I}$, and $\mathcal{J}$ are the dimension of the Hilbert space, the covariance matrix, the classical Fisher information matrix, and the quantum Fisher information matrix, respectively. These inequalities allows one to deduce lower bounds for several metrics of accuracy, such as infidelity or mean square error, considering the impact of the ensemble size. As accuracy metric we employ the infidelity \cite{Fuchs,Wootters} $I(|\tilde\psi\rangle,|\hat\psi\rangle)=1-|\langle\tilde\psi|\hat\psi\rangle|^2$ between an unknown state $|\tilde\psi\rangle$ and its estimate $|\hat\psi\rangle$. The estimation accuracy is given by the expectation or mean value $\bar I(|\tilde\psi\rangle)$ of the infidelity with respect to all possible estimates $|\hat\psi\rangle$ of a fixed unknown state $|\tilde\psi\rangle$, that is,
\begin{equation}
\bar{I}(|\tilde\psi\rangle)=\mathbb{E}[I(|\tilde\psi\rangle,|\hat\psi\rangle)||\hat\psi\rangle]=\int I(|\tilde\psi\rangle,|\hat\psi\rangle)p(|\hat\psi\rangle)d\hat\psi,
\label{MeanInfidelity}
\end{equation}
where $p(|\hat\psi\rangle)$ is the probability density function of obtaining the estimate $|\hat\psi\rangle$. Depending on the general characteristics of the estimation process, the fundamental inequalities lead to various lower bounds for the estimation accuracy. For the estimation of pure states, which have $2(d-1)$ independent parameters, the ultimate quantum estimation accuracy is given by the inequality $\bar{I}(|\tilde\psi\rangle)\ge\bar I_p$, with $\bar I_{p}=(d-1)/N$ \cite{Zhu,Gill-Massar,Hayashi0} the Gill-Massar lower bound. This state-independent bound can be used as a benchmark to assess tomographic methods.

Our main goal is designing a tomographic method for pure quantum states in high dimensions that achieves an estimation accuracy equal to $\bar I_{p}$. The design of the proposed tomographic method originates from observing \cite{Ferrie,Granade,Chapman,Hou} that an unknown and fixed pure quantum state $|\tilde\psi\rangle$ can be characterized as the minimizer of the infidelity in the Hilbert space of the estimate $|\hat\psi\rangle$, that is, $I=0$ when $|\hat\psi\rangle=|\tilde\psi\rangle$. Then, we can envision the use of an optimization method to iteratively drive a sequence of estimates toward decreasing values of the infidelity. 

The choice of the optimization method requires certain consideration. Traditional optimization methods are based on the evaluation of higher derivatives of the function to be optimized. In the case of the infidelity this is not possible, since the derivatives depend on the state to be estimated, which is unknown. We also require a fast convergence rate from the optimization method. Furthermore, the implementation of the method should be as economical as possible from the point of view of computing and physical resources. 

To deal with these situations we resort to the Complex simultaneous perturbation stochastic approximation. This method can optimize real valued functions of complex variables that also depend on unknown complex parameters. The infidelity is a non-holomorphic function, that is, it violates the Cauchy-Schwarz conditions, which are necessary for the existence of a complex derivative. In this case, the usual approach is to optimize with respect to a real parametrization of the complex variables entering in the infidelity, that is, the complex probability amplitudes. This approach has some unwanted side effects. The elements of the real-valued gradient of the infidelity are in general more convoluted than would be those of a complex gradient formed by first order derivatives with respect to the initial complex variables. Additionally, any inherent structures present in the complex derivatives of the infidelity, which could be exploited to enhance the performance of optimization methods, are unused \cite{Sorber}. The CSPSA method has been designed to avoid these unwanted features. The main tool in the formulation of CSPSA is the Wirtinger complex calculus. For a target function  $f({\bm z},{\bm z}^*):\mathbb{C}^n\times\mathbb{C}^n\rightarrow\mathbb{R}$ the Wirtinger derivatives are defined by \cite{Wirtinger}
\begin{equation}
\partial_{z_i}=\frac{1}{2}(\partial_{x_i}-i\partial_{y_i})~{\rm and}~\partial_{z_i^*}=\frac{1}{2}(\partial_{x_i}+i\partial_{y_i}),
\end{equation}
where $x_i$ and $y_i$ are the real and imaginary parts of $z_i$, respectively. These derivatives exist even if $f({\bm z},{\bm z}^*)$ is non-holomorphic. Extremal points of $f({\bm z},{\bm z}^*)$ are completely characterized by the conditions $\partial_{z_i^*}f=0~\forall~i=1,\dots,d$ or, equivalently, $\partial_{z_i}f=0~\forall~i=1,\dots,d$ \cite{Brandwood,Nehari,Remmert}. Thereby, the complex gradient is defined by ${\bm g}=\partial_{{\bm z}^*}f$ with $\partial_{{\bm z}^*}=(\partial_{z_1^*},\dots, \partial_{z_d^*})$. The CSPSA method is defined by the iterative rule \cite{We}
\begin{equation}
\hat{\bm z}_{k+1}=\hat{\bm z}_k-a_k\hat{\bm g}_k(\hat{\bm z}_k,\hat{\bm z}_k^*),
\label{ALGORITHM}
\end{equation} 
where $a_k$ is a positive gain coefficient and $\hat{\bm z}_k$ is the estimate of the minimizer $\tilde{\bm z}$ of $f({\bm z},{\bm z}^*)$ at the k-th iteration. The iteration starts from an initial guess $\hat{\bm z}_0$, which is randomly chosen. Instead of employing the complex gradient ${\bm g}$, CSPSA resorts to an estimator $\hat{\bm g}_k(\hat{\bm z}_k,\hat{\bm z}_k^*)$ for the Wirtinger gradient ${\bm g}$ of $f({\bm z},{\bm z}^*)$ the components of which are defined by
\begin{equation}
\hat{g}_{k,i}=\frac{f(\hat{\bm z}_{k+},\hat{\bm z}_{k+}^*)+\epsilon_{k,+}-(f(\hat{\bm z}_{k-},\hat{\bm z}_{k-}^*)+\epsilon_{k,-})}{2c_k{\Delta}_{k,i}^*},
\label{ESTGRADIENT}
\end{equation}
with $\hat{\bm z}_{k\pm}=\hat{\bm z}_k\pm c_k{\bm\Delta}_k$, $c_k$ a positive gain coefficient, and $\epsilon_{k,\pm}$ describes the presence of noise in the values of $f(\hat{\bm z}_{k\pm},\hat{\bm z}_{k\pm}^*)$. The components of the vector ${\bm \Delta}_k\in\mathbb{C}^n$ are identically and independently distributed random variables in the set $\{\pm1,\pm i\}$. The gain coefficients $a_k$ and $c_k$ control the convergence of CSPSA and are chosen as
\begin{equation}
a_k=\frac{a}{(10k+1+A)^s},~~c_k=\frac{b}{(10k+1)^r},
\end{equation}
where the values of $a, A, s, b$ and $r$ can be adjusted to increase the rate of convergence. The estimates ${\bm {\hat z}}_k$ provided by CSPSA converge asymptotically in mean to the minimizer of $\bm{\tilde z}$ of $f$ and $\hat{\bm g}_k$ is an asymptotically unbiased estimator of the Wirtinger gradient.

The use of an estimator $\hat{\bm g}_k(\hat{\bm z}_k,\hat{\bm z}_k^*)$ for the Wirtinger gradient ${\bm g}$ allows CSPSA to optimize functions with unknown parameters, provided that the values of $f(\hat{\bm z}_{k\pm},\hat{\bm z}_{k\pm}^*)$ can be obtained and the unknown parameters remain constant along the optimization procedure. This is precisely the case of the infidelity. Considering an unknown pure quantum state $|\psi({\bm{\tilde z}})\rangle$ of a qudit given by
\begin{equation}
|\psi({\bm{\tilde z}})\rangle=\frac{1}{\sqrt{N}}\sum_{i=0}^{d-1}\tilde z_i|i\rangle,
\end{equation}
with $N=\sum_{i=0}^{d-1}|\tilde z_i|^2$, and an estimate $|\psi({\bm{\hat z}})\rangle$ of $|\psi({\bm{\tilde z}})\rangle$ given by
\begin{equation}
|\psi({\bm{\hat z}})\rangle=\frac{1}{\sqrt{K}}\sum_{i=0}^{d-1}\hat z_i|i\rangle,
\end{equation}
with $K=\sum_{i=0}^{d-1}|\hat z_i|^2$, the infidelity becomes
\begin{equation}
I|\psi(\bm{\tilde z})\rangle,(|\psi(\bm{\hat z})\rangle)=1-\frac{|{\bm{\hat z}}\cdot\tilde{\bm z}^*|^2}{NK}.
\end{equation}
This function quantifies the deviation of the estimate $|\psi({\bm{\hat z}})\rangle$ from the true unknwon state $|\psi(\tilde{\bm z})\rangle$. In this function $\bm{\hat z}$ is a set of complex variables and $\bm{\tilde z}$ plays the role of a set of fixed unknown complex parameters. According to Eq.\thinspace (\ref{ESTGRADIENT}) CSPSA evaluates at each iteration $k$ the infidelity $I(|\psi(\bm{\tilde z})\rangle),|\psi(\bm{\hat z})\rangle$ at ${\bm{\hat z}}=\hat{\bm z}_{k\pm}$. These values can be obtained by projecting the unknown state $|\psi(\bm{\tilde z})\rangle$ onto a $d$-dimensional orthonormal base $B_{k\pm}=\{|\psi_{i,k\pm}\rangle\}$ (with $i=0,\dots,d-1$) that contains the state $|\psi(\hat{\bm z}_{k\pm})\rangle$. This procedure generates the detection statistics $n_{i,k\pm}$ that lead to the probability distributions $p_{i,k\pm}=n_{i,k\pm}/\sum_jn_{j,k\pm}$, where $N_{est}=\sum_jn_{j,k\pm}$ is the total number of copies of the unknown states employed in the projective measurements. These probability distributions are employed to obtain the estimates
\begin{equation}
I(|\psi(\bm{\tilde z})\rangle,|\psi(\hat{\bm z}_{k\pm})\rangle)=1-p_{0,k\pm}
\end{equation}
for the infidelity, where we have assumed the convention $|\psi_{0,k\pm}\rangle=|\psi(\hat{\bm z}_{k\pm})\rangle$. These estimates together with Eq.\thinspace(\ref{ALGORITHM}) generate the next estimate $|\psi(\hat{\bm z}_{k+1})\rangle$ for the unknown quantum state $|\psi(\tilde{\bm z})\rangle$. The CSPSA method can be understood as a generalization of the Simultaneous perturbation stochastic approach (SPSA) \cite{Spall,Spall1}, a well known stochastic gradient-free optimization method working in the field of the real numbers. SPSA has been proposed \cite{Ferrie} and experimentally demonstrated \cite{Chapman} as tomographic method for pure quantum states. In this context, it has been show \cite{We} that CSPSA achieves a higher convergence rate than SPSA.

\begin{algorithm}
	\caption{CSPSA-MLE tomographic method} 
	\label{Algorithm1}
	\begin{algorithmic}[1]
	    \State Consider an known pure state $|\psi(\tilde{{\bm z}})\rangle$
	    \State Set initial guess $\hat{{\bm z}}_1$, and gain coefficients $a$, $A$, $s$, $b$ and $r$
		\For {$k=1,\ldots, k_{max} $}
		    \State Set $$a_k =\frac{a}{(10k+1+A)^s},\quad c_k = \frac{b}{(10k +1)^r}. $$
            \State Choose $\Delta_{k,i}$ randomly in the set $\{\pm1,\pm i \}$.
            \State Calculate $|\psi_{k\pm}\rangle=|\psi(\hat{{\bm z}}_{k\pm})\rangle$, with $\hat{{\bm z}}_{k\pm} = \hat{{\bm z}}_k\pm c_k{\bm \Delta_k}$.
            \State Choose two bases $B_{k\pm}=\{ |\psi_{i,k\pm}\rangle  \}$ such as $|\psi_{0,k\pm}\rangle= |\psi_{k\pm}\rangle$.
            \State Estimate experimentally the probability distributions $p_{i,k\pm}= |\langle \psi_{i,k\pm}|\psi(\tilde{{\bm z}}) \rangle|^2$ on a sample of size $N_{est}$.
            \State Calculate the Infidelities $I(|\psi(\tilde{{\bm z}})\rangle,|\psi_{k\pm}\rangle)=1-p_{0,k\pm}$.
            \State Estimate the gradient as $$\hat{g}_{k,i} = \frac{I(|\psi(\tilde{{\bm z}})\rangle,|\psi_{k+}\rangle)-I(|\psi(\tilde{{\bm z}})\rangle,|\psi_{k-}\rangle)}{2c_k\Delta_{k,i}^*}. $$
            \State Actualize the guess $\hat{{\bm z}}_{k+1}^0=\hat{{\bm z}}_k - a_k \hat{{\bm g}}_k$.
            \State Refine the guess maximizing the acumulative Likelihood function using $|\psi({\bm z}_{k+1}^0)\rangle$ as starting point,
            $$\hat{{\bm z}}_{k+1} =\arg\max_{|\psi\rangle} L_k(|\psi\rangle), \quad \text{s. t.}\quad \langle \psi|\psi\rangle=1.$$
		\EndFor
	\end{algorithmic} 
\end{algorithm}

CSPSA generates $2$ probability distributions at each iteration, that is, a total of $2d$ different probabilities. However, only two of them are employed to estimate the required values of the infidelity. The remaining $2d-2$ probabilities are not occupied by the algorithm. Thus, CSPSA generates a large amount of accumulated data that is simply discarded. Here, we show that precisely this information can be used to improve the convergence rate of CSPSA in such a way that the estimation accuracy reaches the Gill-Massar lower bound for the estimation of pure quantum states. This is accomplished by resorting to Maximum likelihood estimation, a well known statistical inference method that is extensively employed as a post-processing stage in quantum tomographic methods. MLE is a method in statistical inference aimed at estimating unknown parameters of a population from observed data. The underlying idea is to choose as estimator the maximizer of the probability of obtaining the observed data \cite{Cox,Lehmann}. MLE was introduced in quantum tomography as a post-processing method \cite{Hradil2} to obtain physically acceptable quantum states. A quantum system that undergoes a measurement process described by the set of projectors $\{|a_i\rangle\langle a_i|\}$ has a likelihood function  $L(\rho)$ given by
\begin{equation}
L(\rho)=\frac{N!}{\Pi_j n_j!}\Pi_i Tr(\rho|a_i\rangle\langle a_i|)^{n_i},
\end{equation}
where the detection statistic of each projector is given by the number of counts $n_i$ and the total number of counts is given by $N=\sum_in_i$. If the quantum state prior to the measurement is $\rho$, then $L(\rho)$ is the total joint probability of registering data $\{n_i\}$. MLE is defined by the convex optimization problem
\begin{equation}
\arg\max_{\rho}L(\rho),\, s.\thinspace t.\, Tr(\rho)=1,\, \rho\ge0.
\end{equation}

At this point we link together CSPSA and MLE. Employing the accumulated data $\{n_{i,m\pm}\}$ between iterations $m=1$ until $m=k$ we define the accumulated likelihood $L_k(\rho)$ by the expression
\begin{equation}
L_k(\rho)=\Pi_{m=1}^k\Pi_{\lambda=\pm}\Pi_{i=0}^{d-1}Tr(\rho|\psi_{i,m\lambda}\rangle\langle\psi_{i,m\lambda}|)^{n_{i,m\lambda}},
\end{equation}
which is maximized in the set of pure states employing as starting guess the estimate $|\psi(\hat{\bm z}_{k+1})\rangle$ provided by CSPSA. The refined estimate provided by MLE is then employed as starting guess for the next iteration with CSPSA. We refer to this procedure as the CSPSA-MLE tomographic method. The main steps of the CSPSA-MLE tomographic method have been summarized as pseudocode in Algorithm \ref{Algorithm1} above.

\section*{Results}

For a fixed unknown state, the CSPSA-MLE method exhibits three sources of randomness. Since there is no a priori information about the unknown state, the initial guess is chosen according to a Haar-uniform distribution. At each iteration, the vector ${\bm \Delta}_k$ is also randomly chosen and two measurements are performed on an ensemble of size $N_{est}$, which leads to finite statistics effects. Thus, each time the CSPSA-MLE method is employed to estimate a fixed unknown state $|\psi(\bm{\tilde z})\rangle$, a different estimate $|\psi(\bm{\hat z})\rangle$ is generated. In this scenario, the accuracy of the estimation procedure for a fixed unknown state $|\psi(\bm{\tilde z})\rangle$ is given by the expectation value $\bar I(|\psi(\bm{\tilde z})\rangle)$ of Eq.\thinspace(\ref{MeanInfidelity}).

To study the performance of the CSPSA-MLE method several Monte Carlo experiments in the regime of a small number of iterations were carried out. A set $\Omega_d$ with $2\times10^2$ pure quantum states $|\psi(\bm{\tilde z})\rangle$ of a single $d$-dimensional quantum system (qudit), uniformly distributed on the unit hypersphere, was generated. Each state in $\Omega_d$ was reconstructed via the CSPSA-MLE- method considering a number $G$ of initial guesses, also uniformly distributed on the unit hypersphere, and $R$ independent simulations for each fixed pair of unknown state and initial guess. At each iteration the values of $I(|\psi(\tilde{\bm z})\rangle, |\psi(\hat{\bm z}_{k,\pm})\rangle)$ were estimated considering a multinomial distribution on an ensemble of size $N_{est}$. For each state in $\Omega_d$; mean, variance, median and interquartile range for the infidelity as functions of the number of iterations $k$ for several values of ensemble size $N_{est}$ were estimated. Similar numerical experiments were performed via CSPSA without MLE for comparison purposes. Since the optimization of the gains is a computational costly problem, we have resorted to the gains $s=1$ and $r=0.166$. These lead to a high rate of convergence for CSPSA in the regime of few iterations. The resting coefficients have been set to $A=0, a=3$, and $b=0.35,0.3,0.07,0.06,0.03$ for $N_{est}=10,10^2,10^3,10^4,10^5$, correspondingly.

\begin{figure}[t]
\centerline{\includegraphics[width=\textwidth]{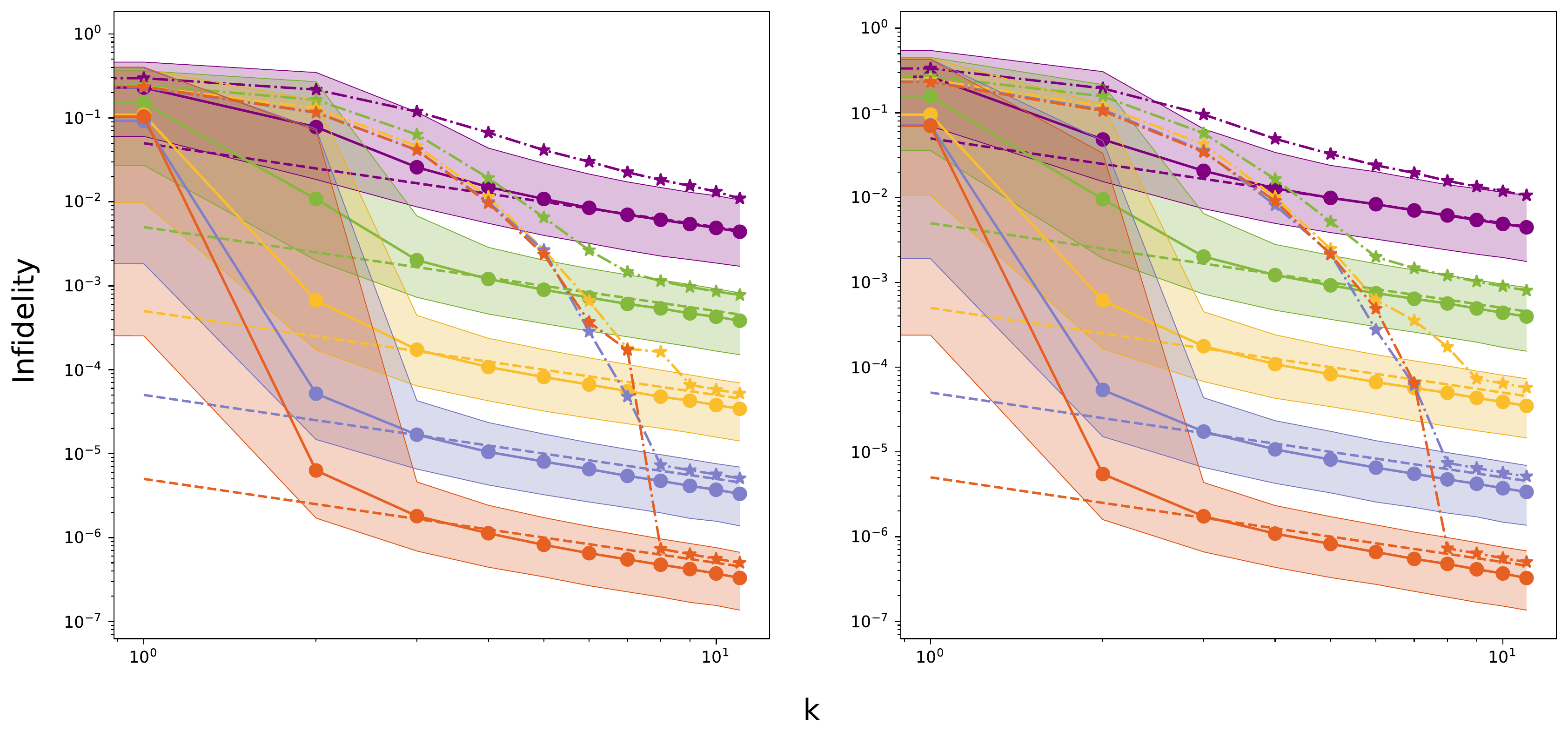}}
\caption{Mean and median infidelity $\bar I$ (stars) and $\bar M$ (circles), respectively, as functions of the number $k$ of iterations achieved by CSPSA-MLE method for two randomly selected pure states of a single qubit in $\Omega_2$ with with $N_{est}=10$ (violet), $10^2$ (green), $10^3$ (yellow), $10^4$ (light blue) and $10^5$ (purple) from top to bottom. Dashed lines indicate the Gill-Massar lower bound $\bar I_p$ for $N=2N_{est}k$. Shaded areas represent interquartile range.  Values of gain coefficients are: $s=1$, $r=1/6$, $a=3$, $A=0$, and $b=0.35, 0.3, 0.07, 0.06, 0.03$ for $N_{est}=10, 10^2, 10^3, 10^4, 10^5$, respectively.}
\label{Fig1}
\end{figure}

Fig.\thinspace\ref{Fig1} displays a log-log graphic of the mean infidelity $\bar I(|\psi(\bm{\tilde z})\rangle)$ (stars), for two randomly chosen states $|\psi(\bm{\tilde z})\rangle$ in $\Omega_2$, that is, for a single qubit, as a function of the number $k$ of iterations and for several values of the ensemble size $N_{est}$. This mean infidelity is estimated as
\begin{equation}
\bar I(|\psi(\bm{\tilde z})\rangle)=\frac{1}{RG}\sum_{\bm{\hat z}}I(|\psi(\bm{\hat z})\rangle,|\psi(\bm{\tilde z})\rangle),
\label{MeanInfidelityEstimate}
\end{equation}
with $G=500$ and $R=20$. Within the first 10 iterations the mean infidelity $\bar I(|\psi(\bm{\tilde z})\rangle)$ exhibits a fast decrease that is followed by an asymptotic linear behavior. The decrease of $\bar I(|\psi(\bm{\tilde z})\rangle)$ becomes more pronounced as $N_{est}$ increases. After 10 iterations the CSPSA-MLE method leads to a mean infidelity $\bar I(|\psi(\bm{\tilde z})\rangle)$ approximately equal to $10^{-2}, 7\times10^{-4}, 5\times10^{-5}, 5\times10^{-6}$ and $5\times10^{-7}$, for increasing $N_{est}$. For the same states CSPSA without MLE yields after 10 iterations a mean infidelity $\bar I(|\psi(\bm{\tilde z})\rangle)$ approximately equal to $2.1\times10^{-1}, 5.2\times10^{-2}, 4.1\times10^{-2}, 4.0\times10^{-2}$ and $3.9\times 10^{-2}$ (see Fig.1 in \cite{We}), for increasing $N_{est}$. Thus, the concatenation of CSPSA to MLE yields a mean infidelity that is $10^{-1}$ to $10^{-5}$ times closer to the true minimum than the one provided by CSPSA alone. Let us note that this comparison considers the same type and amount of physical resources for both methods, that is, ensemble size $N_{est}$ for the estimation of the infidelity and total number $2dk$ of measurement outcomes. The estimation via CSPSA can reach similar mean infidelity values to the CSPSA-MLE method but at the expense of more iterations. For instance, with $N_{est}=10^4$ and after 10 iterations the CSPSA-MLE method delivers a mean infidelity of $5\times10^{-6}$. A similar value can be achieved via estimation with CSPSA with $N_{est}=10^4$ after 100 iterations, which represents an increase of one order of magnitude in the total ensemble size $N$ as well as in the total number of measurements. Thus, the concatenation of MLE to CSPSA provides a significative improvement in the rate of convergence and a large reduction of the required physical resources. 

\begin{figure}[t]
\centerline{\includegraphics[clip,width=\textwidth]{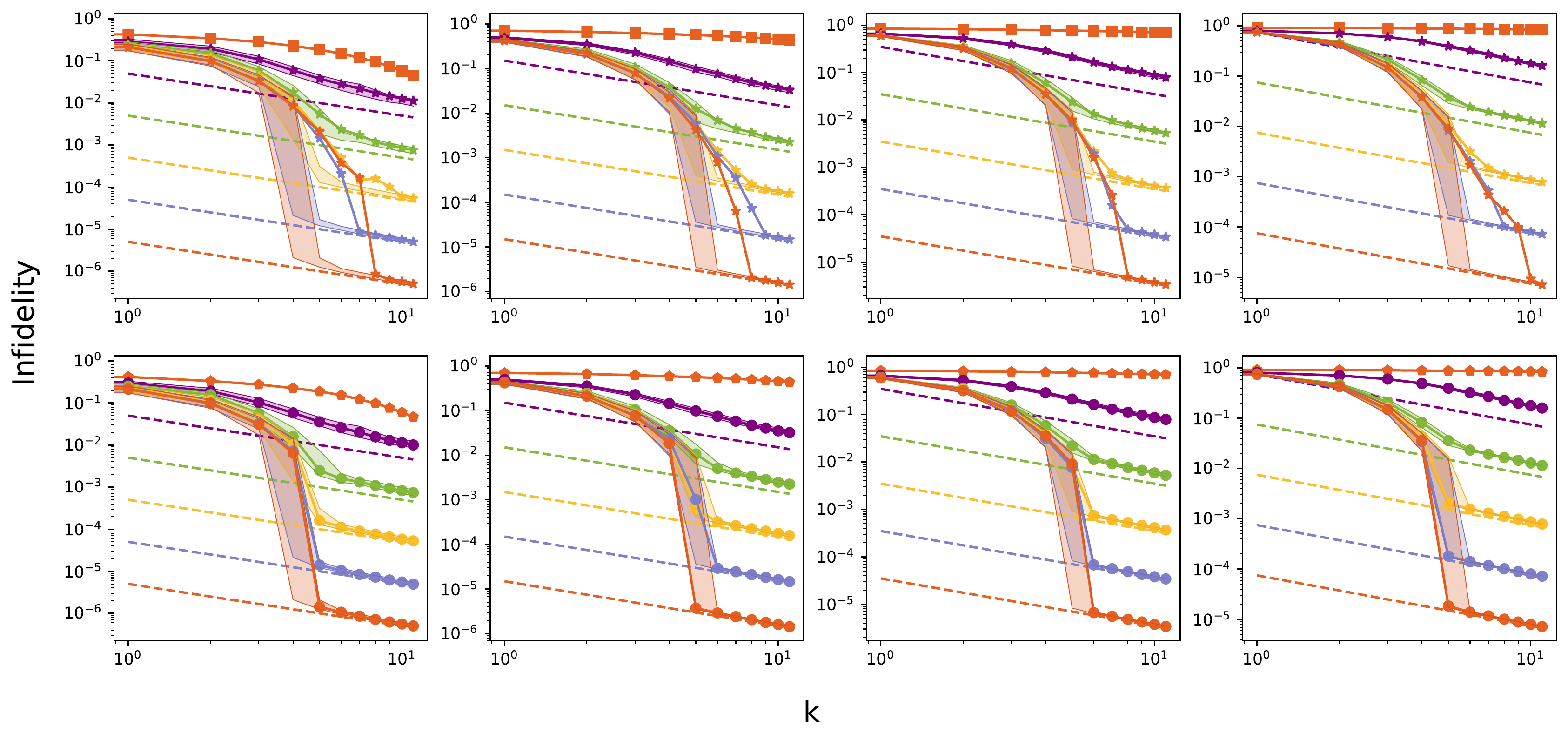}}
\caption{Mean and median infidelity $\bar{\mathbb I}$ (stars, upper row) and $\bar{\mathbb M}$ (solid circles, lower row) as functions of the number $k$ of iterations obtained via the CSPSA-MLE method for the estimation of single $d$-dimensional pure quantum states in $\Omega_d$ for $d=2,4,8,16$, from left to right, with $N_{est}=10$ (violet), $10^2$ (green), $10^3$ (yellow), $10^4$ (light blue) and $10^5$ (purple), from top to bottom. Shaded areas describe interquartile range. Dashed lines indicate the value of the Gill-Massar lower bound $\bar I_p$ for $N=2kN_{est}$. Mean $\bar{\mathbb I}$ (solid orange squares, upper row) and median $\bar{\mathbb M}$ (solid orange rombos, lower row) as functions of the number $k$ of iterations obtained via the CSPSA method for the estimation of single $d$-dimensional pure quantum states in $\Omega_d$ for $d=2,4,8,16$ with $N_{est}=10^5$. Gain coefficients as in Fig.\thinspace\ref{Fig1}.}
\label{Fig2}
\end{figure}

After 7 iterations and for $N_{est}=10,10^2,10^3,10^4,10^5$ the mean infidelity $\bar I(|\psi(\bm{\tilde z})\rangle)$ asymptotically approximates linear behavior. This resembles the Gill-Massar lower limit $\bar I_{p}$ for the average infidelity reached in the estimation of pure states, where now $N=2kN_{est}$ is the total ensamble size employed after $k$ iterations. This bound imposes a fundamental precision limit on the achievable mean infidelity: no method for the estimation of pure states attains a mean infidelity $\bar I(|\psi(\bm{\tilde z})\rangle)$ lower than $\bar I_{p}$. Dashed lines in Fig.\thinspace\ref{Fig1} correspond to $\bar I_{p}$ as function of $k$ for several values of $N_{est}$. Clearly, the CSPSA-MLE method delivers a mean infidelity of the same order of magnitude than $\bar I_{p}$, albeit slightly higher. However, the gap between $\bar I(|\psi(\bm{\tilde z})\rangle)$ and $\bar I_{p}$ tends to close asymptotically as $k$ increases. Thus, the infidelity provided by the CSPSA-MLE method tends to approach the Gill-Massar lower bound $\bar I_{p}$ with a convergence rate that increases with $N_{est}$. 

Fig.\thinspace\ref{Fig1} also displays the median infidelity $\bar{M}(|\psi(\bm{\tilde z})\rangle)$ of  $I(|\psi(\bm{\hat z})\rangle,|\psi(\bm{\tilde z})\rangle)$ as a function of $k$ for both randomly chosen states $\psi(\bm{\hat z})\rangle$ and for several values of $N_{est}$. This exhibits a much faster decrease and an earlier onset of the asymptotic linear behavior. The shaded areas in Fig.\thinspace\ref{Fig1} correspond to the interquartile range, which is divided into two areas above and below the median infidelity $\bar{M}(|\psi(\bm{\tilde z})\rangle)$. In the lineal regime, the Gill-Massar lower bound $\bar I_{p}$ lays in the upper half of the interquartile range. This indicates that more than 50\% of the reconstruction attempts lead to an infidelity lower than $\bar I_{p}$. Nevertheless, the mean infidelity $\bar I(|\psi(\bm{\tilde z})\rangle)$ is above the median $\bar{M}(|\psi(\bm{\tilde z})\rangle)$. This points to the existence of a small fraction of realizations with high values of the infidelity, which increases the value of the mean infidelity above the median infidelity. As the number of iterations increases the impact of these realizations on the value of the mean infidelity decreases and mean and median infidelity tend to reach similar values.

The main features exhibited in Fig.\thinspace\ref{Fig1} are typical, that is, all randomly chosen states in $\Omega_d$ display a similar behavior. This is shown in Fig.\thinspace\ref{Fig2} where mean $\bar{\mathbb I}$ (stars), median $\bar{\mathbb M}$ (dots) and interquartile range (shaded areas) of the mean infidelity $\bar I(|\psi(\bm{\tilde z})\rangle)$ over all states in $\Omega_d$ as a function of $k$ for several values of $N_{est}$ are depicted and compared to $\bar I_p$ (dashed lines). The mean $\bar{\mathbb I}$ (stars) correspond to the expectation value of $\bar I(|\psi(\bm{\tilde z})\rangle)$ on the Hilbert space, that is,
\begin{equation}
\bar{\mathbb I}=\mathbb{E}[\bar I(|\tilde\psi\rangle)||\tilde\psi\rangle]=\int \bar I(|\tilde\psi\rangle)f(|\tilde\psi\rangle)d\tilde\psi,
\end{equation}
where $f(|\tilde\psi\rangle)$ is the probability density function of randomly and uniformly generating the unknown state $|\tilde\psi\rangle$. The mean $\bar{\mathbb I}$ is estimated as
\begin{equation}
\bar{\mathbb I}=\frac{1}{200}\sum_{|\psi(\bm{\tilde z})\rangle\in\Omega}\bar I(|\psi(\bm{\tilde z})\rangle).
\end{equation}

Fig.\thinspace\ref{Fig2} exhibits a fast decrease of $\bar{\mathbb I}$ until approximately iteration 7, which is followed by asymptotic linear behavior. The median $\bar{\mathbb M}$ shows a similar behavior, although it enters an asymptotic linear behavior faster than the mean $\bar{\mathbb I}$ in iteration 5 approximately. As the number $k$ of iterations increases mean $\bar{\mathbb I}$ and median $\bar{\mathbb M}$ overlap almost perfectly in the linear regime. The interquartile range is very narrow and nearly indistinguishable from these two quantities. This indicates that in the linear regime the mean infidelities $\bar I(|\psi(\bm{\tilde z})\rangle)$ of the states in $\Omega_d$ are  concentrated in an extremely narrow interval around the mean $\bar{\mathbb I}$ and the median $\bar{\mathbb M}$. Thus, all states in $\Omega_d$ are estimated by the CSPSA-MLE method with an accuracy that is very close to $\bar{\mathbb I}$. Furthermore, the estimation accuracy $\bar{\mathbb I}$ provided by the CSPSA-MLE method tends to converge from above to $\bar I_{p}$. Thereby, the CSPSA-MLE method produces a mean infidelity $\bar{\mathbb I}$ for any unknown pure state that approaches asymptotically the best possible estimation accuracy of pure states $\bar I_p$ allowed by the laws of quantum mechanics. Fig.\thinspace\ref{Fig2} also displays the behavior of $\bar{\mathbb I}$ (squares) and median $\bar{\mathbb M}$ (rombos) obtained with the use of CSPSA only, that is, without employing MLE to refine the guesses provided by CSPSA, for the case of $N_{est}=10^5$. Clearly, the CSPSA-MLE tomographic method outperforms the CSPSA tomographic method by at least 5 orders of magnitude. Therefore, the concatenation of MLE to CSPSA plays a key role in improving the accuracy of the estimation and bringing it closer to the lower bound of Gill-Massar.

Once the CSPSA-MLE method enters in a lineal regime, after approximately 10 iterations in the inspected dimensions, delivers an estimation accuracy $\bar{I}(|\tilde\psi\rangle)$ close to $\bar I_p$. Thereby, we can write
\begin{equation}
N\approx\frac{d-1}{\bar{I}(|\tilde\psi\rangle)},
\end{equation}
or since $N=2N_{est}k$ also
\begin{equation}
N_{est}k\approx\frac{d-1}{2\bar{I}(|\tilde\psi\rangle)}~~~{\rm for}~k\ge10.
\end{equation}
Thus, in a given dimension $d$ the CSPSA-MLE tomographic method achieves a predefined estimation accuracy with an ensamble size $N$ that can be divided into $2k$ ensambles of size $N_{est}$ each one. Thereby, we can employ a small value of $N_{est}$ and a large number of iterations or a large value of $N_{est}$ and a small number of iterations. The last alternative provides a faster convergence rate. In this case, the estimation of the complex gradient is closer to the exact gradient and the convergence of CSPSA becomes similar to a complex formulation of a deterministic first-order iterative optimization algorithm. However, whether the first or second alternative is more appropriate also depends largely on the characteristics of the experimental platform where the estimation is realized.

We can also compare the mean infidelity $\bar{\mathbb I}$ achieved by the CSPSA-MLE method with the Gill-Massar lower bound $\bar I_{m}$ for the mean infidelity achieved in the estimation of full-rank mixed states via separable measurements on the ensemble of equally prepared copies, which is given by $\bar I_{m}=[(d+1)/2]^2\bar I_{p}$. Tomographic methods designed to estimate unknown mixed states cannot achieve a better accuracy than $\bar I_{m}$ as long as they resort to separable measurements on the ensemble of equally prepared copies. This departs quadratically from $\bar I_p$ and $\bar{\mathbb I}$ as the dimensions increases. Thus, tomographic methods that do not employ the a priori information about the purity of the unknown state cannot estimate pure states with an accuracy better than $\bar I_{m}$. Thereby, the CSPSA-MLE method provides an advantage for pure state estimation over standard quantum tomography, two-stage standard quantum tomographic, and methods such as the ones based on mutually unbiased bases, symmetric informationally complete positive-operator-valued measures, and equidistant states.

\section*{Discussion}

The accurate estimation of quantum states with a limited ensemble size is a difficult task. In the case of 2-dimensional pure quantum states, the best measurement strategy, that is, the one that leads to the (local unbiased) estimator that saturates the quantum Cram\'er-Rao bound, is generally a function of the parameters of the unknown state itself. Thereby, the use of the optimal estimation strategy is unfeasible \cite{Cochran}. It is possible, however, to employ an adaptive strategy that approaches the quantum Cram\'er-Rao bound. This strategy consists of a sequence of measurements where each one is optimal for a given guess of the unknown state \cite{Nagaoka,Okamoto}. The multi-parameter estimation of an unknown $d$-dimensional quantum state, which is defined by $2d-2$ independent real numbers,  cannot be carried out following a similar strategy since the optimal measurement strategy is unknown.

Instead, we have approached the estimation of unknown $d$-dimensional quantum states from the optimization of the metric used to characterize the accuracy of the estimation process. The optimization is solved by a method based on the concatenation of CSPSA to MLE.  The proposed a method allows for estimating quantum pure states with high accuracy. CSPSA drives a sequence of projective measurements toward the infidelity minimizer. MLE provides at each iteration a refinement of the estimates considering the accumulated data generated by all previous measurements. Monte Carlo experiments for dimensions $d=2,4,8$ and $16$ indicate that the mean infidelity for a fixed arbitrary unknown state exhibits a fast decrease within few iterations followed by an asymptotic linear trend. In the linear regime the attained mean infidelity closely approaches the fundamental limit on the accuracy established by the Gill-Massar lower bound for the mean infidelity of the estimation of pure states. Hence, the CSPSA-MLE method surpasses the accuracy of known tomographic methods for pure quantum states. The mean infidelity is also lower than the Gill-Massar lower bound for the infidelity of the estimation of mixed quantum states. Consequently, no tomographic method for mixed states can achieve a mean infidelity lower than the one attained by the CSPSA-MLE method. The median infidelity is also below the Gill-Massar lower bound for pure states. Therefore, more than 50\% of the estimation attempts leads to lower infidelities than the Gill-Massar lower bound. 

The CSPSA-MLE method exhibits a clear trade-off. The concatenation of CSPSA to MLE increases the rate of convergence of the mean infidelity, which leads to a decrease in the number of iterations and in the ensemble size. However, there is an increase in the computational complexity of the algorithm because in each iteration it is now necessary to solve the optimization problem corresponding to MLE. Recently, optimized methods for MLE in high dimension have been proposed \cite{Shang,Bolduc}.

An experimental realization of the CSPSA-MLE method applied to state of a single 2-dimensional quantum system can be carried out with current experimental techniques \cite{Chapman,Okamoto} for generating and measuring single-photon polarization states. The higher dimensional case can be demonstrated by means of experimental setups based on single photons and concatenated spatial light modulators \cite{Goyeneche,Neves,Daniel} or via integrated quantum photonics \cite{Wang,Carine}. These two experimental platforms offer the possibility of performing electronically controlled adaptive measurements.

\section*{Code availability}

The source codes that support the results of this study are available from the corresponding author upon reasonable request.

\section*{Acknowledgements}

This work was funded by the Millennium Institute for Research in Optics and by CONICYT Grant 1180558. L.\thinspace Z. and L.\thinspace P. acknowledge support by CONICYT Grants 22161286 and 22161371, respectively.

\section*{Author contributions statement}

A. D. conceived the stochastic optimization method on the field of complex numbers and  L. Z. proposed the concatenation of statistical inference  to CSPSA. L. Z. and L. P. performed numerical experiments. A. D. and S. N. analyzed the results of numerical experiments. A. D. wrote the article with input from all authors. All authors reviewed and contributed to the final manuscript.

\section*{Additional information}

\textbf{Competing interests:} The authors declare no competing interests.


\begin{thebibliography}{99}

\bibitem{Wootters-Zurek} Wootters, W. K.  \& Zurek, W. H. A single quantum cannot be cloned. \href{https://www.nature.com/articles/299802a0}{{\it Nature} {\bf 299}, 802 (1982).} 

\bibitem{Ivanovic} Ivanovic, I. D. How to differentiate between non-orthogonal states. \href{https://www.sciencedirect.com/science/article/abs/pii/0375960187902222}{{\it Phys. Lett. A} {\bf 123,} 257 (1987).}

\bibitem{Dieks} Dieks, D. Overlap and distinguishability of quantum states. \href{https://www.sciencedirect.com/science/article/abs/pii/0375960188908407}{{\it Phys. Lett. A} {\bf 126,} 303 (1988).}

\bibitem{Peres0} Peres, A. How to differentiate between non-orthogonal states. \href{https://www.sciencedirect.com/science/article/abs/pii/0375960188910341}{{\it Phys. Lett. A} {\bf 128,} 19 (1988).}

\bibitem{vonNeumann} Von Neumann, J. \href{https://press.princeton.edu/books/hardcover/9780691178561/mathematical-foundations-of-quantum-mechanics}{{\it Mathematical Foundations of Quantum Mechanics} (Princeton U. Press, New Jersey, 1983).}

\bibitem{Holevo} Holevo, A.\thinspace S. \href{https://www.springer.com/la/book/9788876423758}{{\it Probabilistic and Statistical Aspects of Quantum Theory} (North-Holland, Amsterdam, 1982).} 

\bibitem{Peres} Peres, A. \href{https://www.springer.com/gp/book/9780792325499}{{\it Quantum Theory: Concepts and Methods} (Kluwer, Dordrecht, 1993).}

\bibitem{Helstrom} Helstrom, C.\thinspace W. \href{https://www.elsevier.com/books/quantum-detection-and-estimation-theory/helstrom/978-0-12-340050-5}{{\it Quantum Detection and Estimation Theory} (Academic Press, New York, 1976), Vol. 84.}

\bibitem{Helstrom2} Helstrom, C.\thinspace W. The minimum variance of estimates in quantum signal detection. \href{https://ieeexplore.ieee.org/document/1054108}{{\it IEEE Transactions on Information Theory} {\bf 14,} 234 (1968).}

\bibitem{Braunstein} Braunstein, S.\thinspace L. \& Caves, C.\thinspace M. Statistical distance and the geometry of quantum states. \href{https://journals.aps.org/prl/abstract/10.1103/PhysRevLett.72.3439}{{\it Phys. Rev. Lett.} {\bf 72,} 3439 (1994).}

\bibitem{Paris} Paris, M. \& \v Reh\'a\v cek, J. eds., \href{http://www.springer.com/la/book/9783540223290}{{\it Quantum State Estimation}, Lecture Notes in Physics, vol. 649 (Springer-Verlag, Berlin Heidelberg, 2004).}

\bibitem{DiVincenzo} DiVincenzo, D.\thinspace P. The physical implementation of quantum computation. \href{https://onlinelibrary.wiley.com/doi/10.1002/1521-3978%28200009%2948%3A9/11%3C771%3A%3AAID-PROP771%3E3.0.CO%3B2-E}{{\it Fortschr. Phys.} {\bf 48,} 771 (2000).}

\bibitem{Kok} Kok, P., Munro, W.\thinspace J., Nemoto, K., Ralph, T.\thinspace C., Dowling, J.\thinspace P. \& Milburn, G.\thinspace J. Linear optical quantum computing with photonic qubits. \href{https://journals.aps.org/rmp/abstract/10.1103/RevModPhys.79.135}{{\it Rev. Mod. Phys.} {\bf 79,} 135 (2007).}

\bibitem{Nielsen}Nielsen, M.\thinspace A. \& Chuang, I. L. \href{https://www.cambridge.org/core/books/quantum-computation-and-quantum-information/01E10196D0A682A6AEFFEA52D53BE9AE}{{\it Quantum Computation and Quantum Information: 10th Anniversary Edition} (Cambridge University Press, New York, 2010).}

\bibitem{Giovannetti} Giovannetti, V. Lloyd, S. \& Maccone, L. Quantum-Enhanced Measurements: Beating the Standard Quantum Limit. \href{http://dx.doi.org/10.1126/science.1104149}{{\it Science} {\bf 306,} 1330 (2004).}

\bibitem{Banaszek} Banaszek, K., Cramer, M. \& Gross, D. Focus on quantum tomography. \href{http://iopscience.iop.org/article/10.1088/1367-2630/15/12/125020}{{\it New J. Phys.} {\bf 15,} 125020 (2013).}

\bibitem{Wootters2} Wootters, W.\thinspace K. \& Fields, B.\thinspace D. Optimal state-determination by mutually unbiased measurements \href{https://www.sciencedirect.com/science/article/pii/0003491689903229}{{\it Ann. Phys. (NY)} {\bf 191,} 363 (1989).}


\bibitem{James} James, D.\thinspace F.\thinspace V., Kwiat, P.\thinspace G., Munro, W.\thinspace J. \& White, A.\thinspace G. Measurement of qubits. \href{https://journals.aps.org/pra/abstract/10.1103/PhysRevA.64.052312}{{\it Phys. Rev. A} {\bf 64,} 052312 (2001).}

\bibitem{Renes} Renes, J.\thinspace M., Blume-Kohout, R., Scott, A.\thinspace J. \& Caves, C.\thinspace M. Symmetric informationally complete quantum measurements. \href{https://aip.scitation.org/doi/abs/10.1063/1.1737053}{{\it J. Math. Phys.} {\bf 45,} 2171 (2004).}


\bibitem{Gross} Gross, D., Liu, Y.-K., Flammia, S. T., Becker, S. \& Eisert, J. Quantum state tomography via compressed sensing. \href{http://dx.doi.org/10.1103/PhysRevLett.105.150401}{{\it Phys. Rev. Lett.} {\bf 105,} 150401 (2010).}

\bibitem{Cramer} Cramer, M. {\it et. al.} Efficient quantum state tomography. \href{http://dx.doi.org/10.1038/ncomms1147}{{\it Nat. Commun.} {\bf 1,} 149 (2010).}

\bibitem{Ahn} Ahn, D. {\it et al.} Adaptive compressive tomography with no a priori information. \href{http://link.aps.org/doi/10.1103/PhysRevLett.122.100404}{{\it Phys. Rev. Lett.} {\bf 122,} 100404 (2019).}

\bibitem{Salazar} Salazar, R. \& Delgado, A. Quantum tomography via unambiguous state discrimination. \href{https://journals.aps.org/pra/abstract/10.1103/PhysRevA.86.012118}{{\it Phys. Rev. A} {\bf 86,} 012118 (2012).}

\bibitem{Paiva} Paiva-Sanchez, C., Burgos-Inostroza, E., Jim\'enez, O. \& Delgado, A. Quantum tomography via equidistant states. \href{https://journals.aps.org/pra/abstract/10.1103/PhysRevA.82.032115}{{\it Phys. Rev. A} {\bf 82,} 032115 (2010).}

\bibitem{Martinez} Mart\'inez, D., Sol\'is-Prosser, M. A., Ca\~nas, G., Jim\'enez, O., Delgado, A. \& G. Lima. Experimental quantum tomography assisted by multiply symmetric states in higher dimensions \href{https://journals.aps.org/pra/abstract/10.1103/PhysRevA.99.012336}{{\it Phys. Rev. A} {\bf 99,} 012336 (2019).}

\bibitem{Mahler} Mahler, D. H. {\it et al.} Adaptive quantum state tomography improves accuracy quadratically. \href{https://journals.aps.org/prl/abstract/10.1103/PhysRevLett.111.183601}{{\it Phys. Rev. Lett.} {\bf 111,} 183601 (2013).}

\bibitem{Guo} Hou, Z., Zhu, H., Xiang, G., Li, C.-F. \& Guo, G.-C. Achieving quantum precision limit in adaptive qubit state tomography. \href{https://www.nature.com/articles/npjqi20161}{{\it npj Quantum Inf.} {\bf 2,} 16001 (2016).}

\bibitem{Pereira} Pereira, L., Zambrano, L., Cort\'es-Vega, J., Niklitschek, S. \& Delgado, A. Adaptive quantum tomography in high dimensions. \href{https://journals.aps.org/pra/abstract/10.1103/PhysRevA.98.012339}{{\it Phys. Rev. A} {\bf 98,} 012339 (2018).}

\bibitem{Gill-Massar} Gill, R.\thinspace D. \& Massar, S. State estimation for large ensembles. \href{https://journals.aps.org/pra/abstract/10.1103/PhysRevA.61.042312}{{\it Phys. Rev. A} {\bf 61,} 042312 (2000).}

\bibitem{We} Utreras-Alarc\'on, A., Rivera-Tapia, M., Niklitschek, S. \& Delgado, A. Stochastic optimization on complex variables and pure-state quantum tomography. \href{https://www.nature.com/articles/s41598-019-52289-0}{{\it Sci. Rep.} {\bf 9}, 16143 (2019).}

\bibitem{Hradil2} Hradil, Z. Quantum-state estimation. \href{http://journals.aps.org/pra/pdf/10.1103/PhysRevA.55.R1561}{{\it Phys. Rev. A
} {\bf 55}, R1561-R1564 (1997).}

\bibitem{Goyeneche} Goyeneche, D. {\it et al.} Five measurement bases determine pure quantum states on any dimension. \href{http://journals.aps.org/prl/abstract/10.1103/PhysRevLett.115.090401}{{\it Phys. Rev. Lett.} {\bf 115,} 090401 (2015).} 

\bibitem{Carmeli} Carmeli, C., Heinosaari, T., Kech M., Schultz, J. \& Toigo, A. Stable pure state quantum tomography from five orthonormal bases. \href{http://iopscience.iop.org/article/10.1209/0295-5075/115/30001/pdf}{{\it EPL} {\bf 115,} 30001 (2016).}

\bibitem{Sosa} Sosa-Martinez, H., Lysne,  N.\thinspace K., Baldwin,  C.\thinspace H., Kalev, A., Deutsch, I.\thinspace H. \& Jessen, P.\thinspace S. Experimental study of optimal measurements for quantum state tomography. \href{https://journals.aps.org/prl/abstract/10.1103/PhysRevLett.119.150401}{{\it Phys. Rev. Lett.} {\bf 119,} 150401 (2017).}

\bibitem{Zambrano} Zambrano, L., Pereira, L. \& Delgado, A. Improved estimation accuracy of the 5-bases-based tomographic method.
\href{https://doi.org/10.1103/PhysRevA.100.022340}{{\it Phys. Rev. A} {\bf 100,} 022340 (2019).}

\bibitem{Paris2} Paris, M. G. A. Quantum estimation for quantum technology. \href{https://www.worldscientific.com/doi/10.1142/S0219749909004839}{{\it Int. J. Quantum. Inform.} {\bf 07,} 125 (2009).}

\bibitem{Cramer-Rao} Helstrom, C.\thinspace W. Minimum mean-squared error of estimates in quantum statistics. \href{https://www.sciencedirect.com/science/article/pii/0375960167903660}{{\it Phys. Lett. A} {\bf 25,} 101 (1967).}

\bibitem{Fuchs} Fuchs, C. \& van de Graaf, J. Cryptographic distinguishability measures for quantum-mechanical states. \href{http://dx.doi.org/10.1109/18.761271}{{\it IEEE Trans. Inf. Theory} {\bf 45,} 1216 (1999).}

\bibitem{Wootters} Wootters, W.\thinspace K. Statistical distance and Hilbert space. \href{https://journals.aps.org/prd/abstract/10.1103/PhysRevD.23.357}{{\it Phys. Rev. D} {\bf 23,} 357 (1981).}

\bibitem{Zhu} Zhu, H. {\it Quantum State Estimation and Symmetric Informationally Complete POMs,} \href{PhD thesis, National Univ. of Singapore (2012).}{http://scholarbank.nus.edu.sg/bitstream/handle/10635/35247/ZhuHJthesis.pdf}

\bibitem{Hayashi0} Hayashi, M. in {\it Quantum Communication, Computing and Measurement} (eds Hirota, O., Holevo, A. S. \& Caves, C. M.) 99–108 (Plenum, 1997).

\bibitem{Ferrie} Ferrie, C. Self-guided quantum tomography. \href{http://journals.aps.org/prl/abstract/10.1103/PhysRevLett.113.190404}{{\it Phys. Rev. Lett.} {\bf 113,} 190404 (2014).}

\bibitem{Granade} Granade, C., Ferrie, C. \& Flammia, S. T. Practical adaptive quantum tomography. \href{https://iopscience.iop.org/article/10.1088/1367-2630/aa8fe6/meta}{{\it New J. Phys.} {\bf 19,} 113017 (2017).} 

\bibitem{Chapman} Chapman, R. J., Ferrie, C. \& Peruzzo, A. Experimental demonstration of self-guided quantum tomography. \href{http://journals.aps.org/prl/pdf/10.1103/PhysRevLett.117.040402}{{\it Phys. Rev. Lett.} {\bf 117,} 040402 (2016).}

\bibitem{Hou} Hou, Z., Tang, J.-F., Ferrie, C., Xiang, G.-Y., Li, C.-F. \& Guo, G.-C. Experimental realization of self-guided quantum process tomography. \href{https://doi.org/10.1103/PhysRevA.101.022317}{{\it Phys. Rev. A} {\bf 101}, 022317 (2020).}

\bibitem{Sorber} Sorber, L., van Barel, M. \& de Lathauwer, L. Unconstrained optimization of real functions in complex variables. \href{http://epubs.siam.org/doi/pdf/10.1137/110832124}{{\it SIAM J. Optim.} {\bf 22,} 879-898 (2012).}

\bibitem{Wirtinger} Wirtinger, W. Zur formalen Theorie der Funktionen von mehr komplexen Veränderlichen. \href{http://link.springer.com/article/10.1007/BF01447872}{{\it Math. Ann.} {\bf 97,} 357 (1927).}

\bibitem{Brandwood} Brandwood, D. H. A complex gradient operator and its application in adaptive array theory. \href{http://ieeexplore.ieee.org/document/4645581/}{{\it Proc. IEE-H} {\bf 130,} 11-16 (1983).}

\bibitem{Nehari} Nehari, Z. {\it Introduction to Complex Analysis} (Allyn \& Bacon, Boston, 1961).

\bibitem{Remmert} Remmert, R. \href{http://www.springer.com/la/book/9780387971957}{{\it Theory of Complex Functions} (Springer-Verlag, New York, 1991).}

\bibitem{Spall} Spall, J.\thinspace C. \href{http://www.wiley.com/WileyCDA/WileyTitle/productCd-0471330523.html}{{\it Introduction to Stochastic Search and Optimization} (Hoboken New Jersey, Wiley, 2003).}

\bibitem{Spall1} Spall, J.\thinspace C. Multivariate stochastic approximation using a simultaneous perturbation gradient approximation. \href{http://ieeexplore.ieee.org/xpl/articleDetails.jsp?arnumber=119632}{{\it IEEE Trans. Autom. Control} {\bf 37,} 332-341 (1992).}

\bibitem{Cox} Cox, D. R. \href{https://www.cambridge.org/core/books/principles-of-statistical-inference/BCD3734047D403DF5352EA58F41D3181#fndtn-information}{{\it Principles of Statistical Inference} (Cambridge University Press, New York, 2006).}

\bibitem{Lehmann} Lehmann, E. \& Casella, G. \href{https://www.springer.com/gp/book/9780387985022}{{\it Theory of Point Estimation} (Springer-Verlag, New York, 1998).}

\bibitem{Cochran} Cochran, W. G. Experiments for Nonlinear Functions. \href{http://dx.doi.org/10.2307/2284499}{{\it J. Am. Stat. Assoc.} {\bf 68,} 771 (1973).}

\bibitem{Nagaoka} Nagaoka, H. in {\it Asymptotic Theory of Quantum Statistical Inference,} edited by M. Hayashi (World Scientific, Singapore, 2005).

\bibitem{Okamoto} Okamoto, R., Iefuji, M., Oyama, S., Yamagata, K., Imai, H., Fujiwara, A. \& Takeuchi, S. Experimental Demonstration of Adaptive Quantum State Estimation. \href{http://dx.doi.org/10.1103/PhysRevLett.109.130404}{{\it Phys. Rev. Lett.} {\bf 109,} 130404 (2012).}

\bibitem{Shang} Shang, J., Zhang, Z. \& Ng, H.\thinspace K. Superfast maximum-likelihood reconstruction for quantum tomography. \href{https://journals.aps.org/pra/abstract/10.1103/PhysRevA.95.062336}{{\it Phys. Rev. A} {\bf 95,} 062336 (2017).}

\bibitem{Bolduc} Bolduc, E., Knee, G.\thinspace C., Gauger, E.\thinspace M. \& Leach, J. Projected gradient descent algorithms for quantum state tomography. \href{http://www.nature.com/articles/s41534-017-0043-1}{{\it npj Quantum Inf.} {\bf 3,} 44 (2017).}

\bibitem{Neves} Sol\'is-Prosser, M. A., Fernandes M. F., Jim\'enez O., Delgado A. \& Neves L. Experimental minimum-error quantum-state discrimination in high dimensions. \href{https://journals.aps.org/prl/abstract/10.1103/PhysRevLett.118.100501}{{\it Phys. Rev. Lett.} {\bf 118,} 100501 (2017).}

\bibitem{Daniel} Mart\'inez, D. {\it et al.} Experimental quantum tomography assisted by multiply symmetric states in higher dimensions. \href{https://journals.aps.org/pra/abstract/10.1103/PhysRevA.99.012336}{{\it Phys. Rev. A} {\bf 99,} 012336 (2019).}

\bibitem{Wang} Wang, J. Sciarrino, F. Laing, A. \& Thompson, M. G. Integrated photonic quantum technologies. \href{https://doi.org/10.1038/s41566-019-0532-1}{{\it Nat. Photon.} {\bf 14,} 273 (2019).}

\bibitem{Carine} Cari\~ne, J., Ca\~nas, G., Skrzypczyk, P.,  \v Supi\'c, I., Guerrero, N., Garcia, T., Pereira, L., Prosser, M. A. S., Xavier, G. B., Delgado, A., Walborn, S. P., Cavalcanti, D. \& Lima, G. Multi-core fiber integrated multi-port beam splitters for quantum information processing. \href{https://www.osapublishing.org/optica/abstract.cfm?uri=optica-7-5-542&origin=search}{{\it Optica} {\bf 7,} 542 (2020).}

\end{thebibliography}
\end{document}